\documentclass[%
 reprint,
 amsmath,amssymb,
 aps,
prb
]{revtex4-2}

\usepackage{graphicx}
\usepackage{dcolumn}
\usepackage{bm}
\usepackage{amsfonts}
\usepackage{color}
\begin{document}

\preprint{APS/123-QED}

\title{
% \color{blue}
Role of conservation laws in %the efficiency of 
the Density Matrix Renormalization Group
%Assessing the use of quantum numbers in infinite gapless DMRG (Need new title)
}

\author{Thomas G. Kiely}
 \altaffiliation{tgk37@cornell.edu}
\author{Erich J. Mueller}
 \altaffiliation{em256@cornell.edu}
\affiliation{Laboratory of Atomic and Solid State Physics, Cornell University, Ithaca, NY 14853}

\date{\today}

\begin{abstract}
We explore matrix product state approximations to wavefunctions which have {\em spontaneously broken} symmetries or are {\em critical}. We are motivated by the fact that symmetries, and their associated conservation laws, lead to block-sparse matrix product states.  
{%\color{blue}
Numerical calculations which 
take advantage of these symmetries run faster and require less memory.
%can often be used to speed-up numerical calculations. 
However, in symmetry-broken and critical phases the
block sparse ansatz yields less accurate energies.
%(unless the bond dimension is increased).
%ground state energies 
%within 
%symmetry-broken and critical phases
% block-sparse matrix product states may provide significantly less accurate ground state energies .  On the other hand, for fixed bond dimension, numerical computations using the block sparse matrices take less memory and time.
%Thus, 
%there is a trade-off between the accuracy of the ansatz and the efficiency of 
%For the symmetry-broken and critical phases studied here, we find that (for a fixed accuracy) block-sparse matrix product states can require a significantly larger bond dimension than dense states that explicitly break the underlying symmetry. 
We characterize the role of conservation laws in matrix product states and determine when
%epending on parameters, 
it is beneficial to make use of them.
%can either be beneficial or detrimental to the memory cost and computer run-time.
% improve or hurt the memory cost and computer run-time.
%Thus, for some purposes, we find that block-sparse matrix product states can be less efficient than their dense counterparts.
}
\end{abstract}

\maketitle

\section{\label{sec:intro}Introduction}

Our most powerful numerical techniques for studying one dimensional quantum systems, such as the Density Matrix Renormalization Group (DMRG) or its time-dependent generalizations \cite{SCHOLLWOCKreview,PAECKEL2019}, are based upon a systematic truncation of the entanglement between neighboring regions of space.
%take advantage of systematic methods of quantifying the entanglement between different regions of space. 
Within these approaches, the quantum mechanical wavefunction has the structure of a matrix product state (MPS): The amplitude of any given configuration is calculated by taking the product of a series of matrices -- one for each site.  
%There are powerful techniques to variationally optimize the matrices~\cite{SCHOLLWOCKreview}, or use them to approximate time dynamics~\cite{PAECKEL2019}.  
In the presence of symmetries, these matrices can be taken to be {\em block-sparse}, where 
%conservation laws force a large number of 
the majority of
matrix elements vanish.  This structure is used in all modern codes to accelerate performance.  Here we assess the limitations of this block-sparse structure: What happens when the symmetry is spontaneously broken, or if one is at a critical point?
%ask what happens when the symmetry is spontaneously broken, or if one is at a critical point.
{%\color{blue} 
Using the transverse-field Ising model as a pedagogical example, we elucidate how 
% in the symmetry broken phase 
the most efficient description of a state 
in the symmetry-broken phase 
does not make use of conservation laws.}
%when the symmetry is spontaneously broken.
%We give a simple argument that if the symmetry is spontaneously broken then the most efficient description of the state will not make use of the conservation laws. 
%Our most important results involve 
%We also 
{%\color{blue} 
We then}
explore
critical systems:
%and we find 
In the superfluid phase the of 1D Bose-Hubbard model and the metallic phase of the 1D Fermi Hubbard model, we find that an MPS which respects the symmetry requires larger matrices to achieve the same accuracy. {%\color{blue}  
For some parameter ranges, this results in a larger memory footprint and longer run-time.}

According to Noether's theorem, 
symmetries are closely related to conservation laws~\cite{Noether1918,noetherEnglish}.
%a system that has a continuous symmetry will conserve some ``Noether charge"~\cite{Noether1918,noetherEnglish}. 
As a relevant example, consider a Hamiltonian that is invariant under the transformation
\begin{equation}
    \hat H\to\hat U\hat H\hat U^\dagger,
    \label{eq:inv}
\end{equation}
where $\hat U(\theta)=e^{i\theta \hat N}$ and $\hat N$ is the total particle number operator. Equation~(\ref{eq:inv}) %implies that $\hat H$ 
%has 
defines a continuous ${\rm U}(1)$ symmetry, parameterized by $\theta$.
%which can intuitively be understood to imply that the phases of single-particle wavefunctions are not observable. The conserved Noether charge implied by Eq.~(\ref{eq:inv}) is the eigenvalue of the operator $\hat N$. More generally,
%Furthermore, Eq.~(\ref{eq:inv}) 
This
can only be true if $[\hat H,\hat N]=0$, which is the formal quantum-mechanical statement that $\hat N$ is conserved.  Consequently, we can find simultaneous eigenstates of $\hat H$ and $\hat N$.

One of the most profound features of many-body physics is that, in the thermodynamic limit, the symmetry may be {\em spontaneously broken} \cite{Beekman2019}:  An infinitesmal symmetry breaking field leads to a ground state which is neither invariant under the symmetry operation, nor is it an eigenstate of the conserved charge.  A relevant example is a  Bose-Einstein condensate, which chooses a particular phase and contains an indefinite number of particles. 
%An infinitesmal symmetry breaking field couples the energy eigenstates from different particle-number blocks.
Spontaneous symmetry breaking is always associated with a ground state degeneracy, and one can restore the symmetry by taking an appropriate quantum superposition of the degenerate ground states.   In the case of a discrete symmetry, such symmetry-restored states are ``Schrodinger cats."  

In Sec.~\ref{sec:TFIsing} we present the transverse-field Ising model, which possesses a discrete $\mathbb{Z}_2$ symmetry, {%\color{blue}
as a pedagogical example}. It has two zero-temperature phases:
% both 
a paramagnetic phase, in which the ground state respects the symmetry; and a ferromagnetic phase, which breaks it.
In both phases one can use a symmetry-preserving MPS to describe the ground state.  In the ferromagnetic phase, however, the resulting MPS corresponds to the aforementioned Schrodinger cat, which is a superposition of the two symmetry-broken solutions.  These symmetry-broken constituents are less entangled than the symmetry-preserving Schrodinger cat, and hence are more efficient to express as a MPS~\cite{SCHOLLWOCKreview}.  Aspects of this behavior are known by the community, but rarely discussed in the literature.

The situation is far more complicated for continuous symmetries.
One-dimensional systems with short-ranged interactions {%\color{blue}
and finite susceptibilities} cannot break a continuous symmetry~\cite{merminWagner,hohenbergLRO,momoi1996}. Instead, strong quantum fluctuations lead to correlation functions that fall off as a power-law~\cite{giamarchi}.  It is far from obvious if they are better described by an MPS that respects the symmetry or one that explicitly breaks it.  We consider two examples:  The Bose-Hubbard model and the Fermi-Hubbard model.
{%\color{blue}
In both cases, we find that the symmetry-conserving MPS requires a larger bond dimension (the linear size of the MPS matrices) to achieve the same accuracy as the symmetry-broken MPS.
}
% {\color{red}
% In both cases we find that for fixed bond dimension (the linear size of the MPS matrices) a symmetry-broken MPS is a better variational wavefunction than one which obeys the  symmetry.  }
For the Fermi-Hubbard case the improvement is fairly modest, while for the Bose-Hubbard case it is quite substantial.  We show that the key difference is the scaling of density fluctuations, which is characterized by the Luttinger parameter $K$, and we quantify this relationship.

{%\color{blue}
Throughout this paper we largely compare sparse and dense MPS representations of states with the {\em same bond dimension}.  This allows us to cleanly understand the ways in which conservation laws manifest in critical and symmetry broken states.  We emphasize that the sparse MPS states with bond dimension $\chi$ are a subset of the dense states with the same bond dimension.  Thus, imposing the symmetry can never improve the energy of the variational ground state. The important question is the {\em extent} to which the dense MPS is a better variational ansatz.

For practical numerical calculations one is likely interested in a different question:  For a fixed numerical accuracy, does it take more computer time to calculate the ground state using a sparse or dense MPS?  One could similarly ask about memory or disk usage. Unfortunately, these questions will inevitably depend on details of the implementation:  How does one store the block-sparse matrices? How does one implement basic linear algebra operations?  We largely relegate these practical questions to Appendix~\ref{sec:efficiency}.  We find that within the ITensor package~\cite{itensor}, runtimes can be either increased or reduced by imposing conservation laws, depending on parameters.  One would intuitively guess that proximity to symmetry breaking would determine the extent to which one benefits from the block-sparse structure.  For the systems we study the Luttinger parameter $K$ quantifies this proximity:  The ideal Bose gas, with $K=0$, has off-diagonal long-range order and is often interpreted as a symmetry-broken state~\cite{pitaevskii2016}. As expected from this argument, the benefits from using dense tensors are largest at small $K$. The other relevant parameter is the target accuracy, which determines the bond dimension.  For low accuracy (small $\chi$), dense calculations are faster, while for high accuracy, block sparse calculations are faster. 
The crossover point depends on $K$:  smaller $K$ favors the dense ansatz.
}

In Section~\ref{sec:ConsLaws} we
{%\color{blue}
review features of the MPS ansatz and discuss}
how Abelian symmetries lead to block-sparse MPS tensors. 
%In Sec.~\ref{sec:Results} we present our results for the transverse-field Ising model (Sec.~\ref{sec:TFIsing}), the Bose-Hubbard model (Sec.~\ref{sec:BoseHubbard}), and the Fermi-Hubbard model (Sec.~\ref{sec:FermiHubbard}).
{%\color{blue}
In Sec.~\ref{sec:Results} we present our results: we begin with the transverse-field Ising model as a pedagogical example (Sec.~\ref{sec:TFIsing}), then move to the more nuanced Bose-Hubbard (Sec.~\ref{sec:BoseHubbard}) and Fermi-Hubbard (Sec.~\ref{sec:FermiHubbard}) models.
}
In Sec.~\ref{sec:Disc} we present a more general interpretation of the results in Secs.~\ref{sec:BoseHubbard} and~\ref{sec:FermiHubbard} in terms of the Luttinger parameter. We conclude in Sec.~\ref{sec:conclusion}.

\section{\label{sec:ConsLaws}Conservation laws in MPS}

An MPS incorporates conservation laws by placing restrictions on which matrix elements can be non-zero~\cite{McCulloch2007}. This sparse structure can be exploited to dramatically speed up tensor contractions. For the purpose of this paper, we will only consider Abelian symmetries generated by a global operator $\hat Q=\sum_i\hat Q_i$ that commutes with the Hamiltonian: $[\hat H,\hat Q]=0$.  
Here $\hat Q_i$ are a set 
of mutually-commuting single-site operators, where $i$ indexes the sites of the MPS in real space -- for concreteness, one can envision $\hat Q=\hat N$, as in Eq.~(\ref{eq:inv}), and take $\hat Q_i=\hat N_i$ to be the number of particles on site $i$.
 %They are requ us to express the local tensors in the eigenbasis of $\hat Q_i$.  
% . This allows us to express the local tensors in the eigenbasis of $\hat Q_i$ (and excludes, e.g. momentum conservation). For concreteness, we will take $\hat Q_i=\hat N_i$ to be the number operator in the example below.
%In general, this encompasses the majority of implemented conservation laws in iDMRG simulations.

%As we will argue, we anticipate our analysis will be valid for non-Abelian symmetries as well.

%In the remainder of this section, we will introduce an implementation of particle number conservation in a finite-sized MPS. We will then consider the ramifications of an infinite system size.

%\subsection{Implementing number conservation}

We consider a matrix product state wavefunction on $L$ sites.
%, composed of rank-3 tensors ${\bf A}_i$ -- here $i$ labels the site, and is not thought of as a tensor index.
The MPS ansatz can be schematically written as
%One schematically writes
$|\psi\rangle =\sum_\sigma  A^{\sigma_1}A^{\sigma_2}\cdots A^{\sigma_L}|\sigma_1\sigma_2\cdots\sigma_L\rangle$ where
$A^{\sigma_i}$ corresponds to a matrix with elements $(A_i)^{\sigma_i}_{s_{i-1}s_i}$. The sum is taken over all $\sigma_1,\sigma_2,\cdots \sigma_L$, where $\sigma_i$ corresponds to the allowed states on site $i$, and over shared indices $s_i$ between adjacent matrices.
%, and the sum is over all $\sigma_1,\sigma_2,\cdots \sigma_L$ where $\sigma_i$ corresponds to the allowed states on site $i$.
%consider an MPS in which the bond dimension $\chi$ is constant throughout. 
%The components of the tensor on site $i$ are %written $A^{\sigma_i}_{s_{i-1},s_i}$ where the %physical index $\sigma_i$ ranges over the %allowed states on site $i$,  
Here $s_{i-1}$ and $s_i$ are the left and right MPS bond indices. The bond dimension $\chi$ is the number of different possible values of $s$.
%One schematically writes
%$|\psi\rangle =\sum_\sigma A^{\sigma_1}A^{\sigma_2}\cdots A^{\sigma_L}|\sigma_1\sigma_2\cdots\sigma_L\rangle$
%Without loss of generality, we will take the bond dimensions (the dimensions of the bond indices) to be constant throughout: ${\rm dim}(s_i)=\chi~\forall~i$. 
In order to make use of the conservation law we write the local Hilbert space in the eigenbasis of the local operator $\hat Q_i$, and define a function $q(\sigma_i)$ which associates a {\em charge} with each of the local basis states: $\hat Q_i |\sigma_i\rangle= q(\sigma_i) |\sigma_i\rangle$. We similarly associate a charge with each possible value of the bond indices. The conservation law is imposed by requiring that the only non-zero elements of $A_i$ obey
% . Hence the values taken by the physical index, $n_i$, should uniquely specify the number of particles on site $i$. The values taken by these indices will be grouped into sets that are identified by distinct quantum numbers, $\{ q(b_i) \}\in \mathbb{Z}$, where $b_i$ is any of the indices of ${\bf A}_i$: $b_i\in\{s_{i-1},n_i,s_i\}$. 
% There are a variety of possible parameterizations one can take for the $\{ q(b_i) \}$. It is instructive to start with the simplest such parameterization: $q(n_i)$ is equal to the number of particles on site $i$.
% %~\footnote{This will not be a convenient parameterization when taking $L\to\infty$, of course, as the $q(s_i)$ will diverge. In that limit, it is convenient to define $q(n_i)$ as the number of particles on site $i$ minus the average density of particles, $\bar n$.}. 
% Requiring that the entire MPS has a definite particle number is equivalent to requiring that the individual MPS tensors ${\bf A}_i$ are block-sparse: the only matrix elements allowed to be non-zero are those that satisfy
\begin{equation}
    q(s_{i-1})+q(\sigma_i)-q(s_i)=0.
    \label{eq:sparsity}
\end{equation}
In the case of number conservation, one can interpret $q(s_{i-1})$  as the number of  particles to the left of site $i$, and  $q(s_{i})$ as the number to the left of site $i+1$.
%With this particular parameterization, one can think of the quantum numbers of the MPS bond indices, $q(s_i)$, as simply the number of particles to the left of site $i$, inclusive, up to some overall constant shift. In cases where the length of the chain diverges, 
For infinite chains, it is convenient to define $q(\sigma_i)$ as the deviation of the quantum number from its average so that the charges of the bond indices are more readily truncated.
%number of particles on a site from the average density, $\bar n$, so that the bond quantum numbers remain finite.

% As a practical example, let's consider a product state: the bond dimension $\chi=1$, and hence each bond index $s_i$ has just a single quantum number $q(s_i)$. Let's define the quantum numbers such that $q(s_{i-1})=N_i$ if there are $N_i$ particles to the left of site $i$~\footnote{Note that this is just one parameterization for the values taken by the quantum numbers -- more often, one implements a subtraction scheme so that the $q(s_i)$ don't grow linearly with system size.}. If there are two particles on site $i$, this parameterization implies that $q(n_i=2)=2$. Hence, one would find
% \begin{equation}
%     q(s_i)=q(s_{i-1})+q(n_i)=N_i+2,
% \end{equation}
% telling us that there are $N_i+2$ particles to the left of site $i+1$.

%To interpret the condition in Eq.~(\ref{eq:sparsity}), recall that a matrix product state is simply a superposition of product states whose coefficients are given by the product of matrices on each site. What 
%Equation~(\ref{eq:sparsity}) mandates that the product states comprising the MPS all have a definite particle number on each site -- that way, given the number of particles left of site $i-1$, $q(n_i)$ uniquely determines $q(s_i)$. As a counterexample, consider the

As should be clear, the block-sparse condition in Eq.~(\ref{eq:sparsity}) greatly reduces the number of matrix elements which need to be stored and speeds up all matrix operations.  Its limitations, however, are illustrated by considering 
a simple Gutzwiller mean-field wavefunction: $|\Psi\rangle=\prod_{i=1}^{L}\otimes \left(a|0\rangle_i+b|1\rangle_i\right)$, which represents a Bose-Einstein condensate in which each site contains the superposition of $0$ and $1$ particle.  
This is a MPS with bond dimension $\chi=1$, but it does not obey Eq.~(\ref{eq:sparsity}). 
One can rewrite it using the conservation laws, but that comes at the cost of greatly increasing $\chi$. For a chain of length $L$, for example, one needs $\chi=L$, and the MPS matrices can take the form
\begin{equation}
A_i=\left(
\begin{array}{cccccc}
a|0\rangle_i&b|1\rangle_i\\
&a|0\rangle_i&b|1\rangle_i\\
&&a|0\rangle_i&b|1\rangle_i\\
&&&\ddots&\ddots
\end{array}
\right).
\label{eq:numConsGutz}
\end{equation}
The rows correspond to configurations where there are $0,1,2,\cdots$ particles to the left of this site.
The bond index increments whenever a site is occupied.

\section{\label{sec:Results}Results}

We characterize the distinction between a quantum-number-conserving (sparse) MPS and a non-conserving (dense) MPS by running iDMRG simulations on a few well-known models. We begin
{%\color{blue}
with a pedagogical discussion of}
%by discussing 
the transverse-field Ising in Sec.~\ref{sec:TFIsing}, which exhibits discrete spontaneous symmetry breaking. In this particular model,
{%\color{blue} 
which has been studied extensively with a wide range of analytical and numerical techniques~\cite{pfeuty1970,pfeuty1971,coleman,schollwock2005},}
we show how one can explicitly construct the dense MPS out of the sparse MPS (and vice versa). 
{%\color{blue}
This transformation preserves the variational energy but not the bond dimension, and hence yields insight into the relative efficiency of the dense and sparse ansatze.}
We then move to examples of Luttinger liquids, namely the Bose-Hubbard (Sec.~\ref{sec:BoseHubbard}) and Fermi-Hubbard (Sec.~\ref{sec:FermiHubbard}) models. These are more complicated systems that do not explicitly break any symmetries, so they necessitate more detailed numerical comparisons.
% {\color{blue} In particular, as they are gapless systems, matrix product states at finite bond dimension cannot represent important qualitative features of the true ground state. For this reason, there is no simple mapping between dense and sparse ansatze. For the purpose of providing consistent and replicable results, we will define ``efficiency" as the accuracy of a ground state wavefunction with fixed bond dimension $\chi$. We provide a discussion of alternative definitions in Appendix...}
We make use of the ITensor library for an efficient implementation of quantum number conservation~\cite{itensor}. Further details of the numerical simulations are discussed in Appendix~\ref{sec:details}.

%We characterize the distinction between a dense MPS and a number-conserving (sparse) MPS by running iDMRG simulations on two well-known strongly-interacting lattice models: the Bose-Hubbard model and the Fermi-Hubbard model. Simulations are performed for relatively low bond dimensions, $\chi=40-120$, with and without number conservation. We make use of the ITensor library~\cite{itensor}. For cases described here, both models have gapless ground states and an emphasis is placed on finite-entanglement scaling~\cite{pollmann2009,Kiely2022} to extract properties of the low-energy effective theories.

\subsection{\label{sec:TFIsing}Transverse-field Ising model: a pedagogical example}

The one-dimensional transverse-field Ising model is an exactly-solvable model of $s=1/2$ spins on a lattice with nearest-neighbor interactions. The Hamiltonian is given by
\begin{equation}\label{tfiham}
    \mathcal{H}_{\rm TFI}=-J\sum_j\big(\sigma^x_j \sigma^x_{j+1}+\alpha~\sigma^z_j\big)
\end{equation}
where $\sigma_j^\beta$ is the Pauli spin matrix ($\beta=x,y,z$) acting on the spin on site $j$. The ratio of the transverse field strength to the nearest-neighbor interaction strength, $\alpha$, is the only non-trivial parameter in the ground state phase diagram (here we consider the ferromagnetic model: $J,\alpha>0$). While $\mathcal{H}_{\rm TFI}$ does not conserve total magnetization, it has a global $\mathbb{Z}_2$ symmetry, $[\mathcal{H}_{\rm TFI},\hat P]=0$, where  the parity operator, $\hat P = \prod_j \sigma^z_j=\exp(i\pi \sum_j (\sigma^z_j-1)/2)$, rotates all spins about the $\hat z$ axis by $\pi$.
%is evident from the fact that $[\mathcal{H}_{\rm TFI},\hat P]=0$ where $\hat P = \prod_j \sigma^z_j$ is the parity operator and $\prod_j$ runs over all sites of the 1D lattice. 
This parity symmetry implies that the magnetization along the $\hat z$ direction is conserved modulo 2. 

The transverse-field Ising model has two zero-temperature phases. When $\alpha>1$, the $\sigma^z$ term dominates and spins tend to align with the transverse field. This phase is even under parity transformations: $\hat P|\psi_{\alpha>1}\rangle=|\psi_{\alpha>1}\rangle$. When $\alpha<1$, the exchange term dominates and spins will tend to align with one another in the $\pm \hat x$ direction. In the thermodynamic limit, an infinitesmal field in the $\hat x$ direction will result in a ground state $|+\rangle$ with a finite magnetization in the 
$+\hat x$ direction.  This  is an example of spontaneous symmetry breaking:  The state 
$|-\rangle =\hat P|+\rangle$ is orthogonal to $|+\rangle$ and has a magnetization in the $-\hat x$ direction.  
Two parity-conserving ground states can be formed by taking $|+\rangle \pm |-\rangle$.
%As spins in the $\hat x$ direction are flipped to the $-\hat x$ direction under parity transformations, the system will tend to spontaneously break parity symmetry in this phase. A parity-conserving ansatz captures the definite-parity eigenstates of the system, which will generically take the form of a macroscopic cat state: $|\psi^{(1)}_{sb}\rangle\pm\hat P|\psi^{(2)}_{sb}\rangle$, where $|\psi^{(i)}_{sb}\rangle$ are the two degenerate symmetry-broken states. 
In the limit $\alpha\to 0$, parity-conserving ground states are given by GHZ states~\cite{GHZ},
\begin{equation}
    %|\psi_{\alpha=0}^\pm\rangle=
    \frac{1}{\sqrt{2}}\big(|\rightarrow\rightarrow\rightarrow\ldots\rangle\pm|\leftarrow\leftarrow\leftarrow\ldots\rangle\big),
\end{equation}
where $|\rightarrow\rangle_j$ denotes a spin on site $j$ oriented in the $\hat x$ direction.
%In the thermodynamic limit, the energy splitting between $|\psi_{\alpha=0}^\pm\rangle$ and the product states $|\leftarrow\leftarrow\leftarrow\ldots\rangle$ and $|\rightarrow\rightarrow\rightarrow\ldots\rangle$ vanishes. Hence the model will often choose the broken-parity states, as they have a lower entanglement entropy. For this reason, the $\alpha<1$ phase is often referred to as the symmetry-broken phase.

\begin{figure}
    \centering
    \includegraphics[width=3.375in]{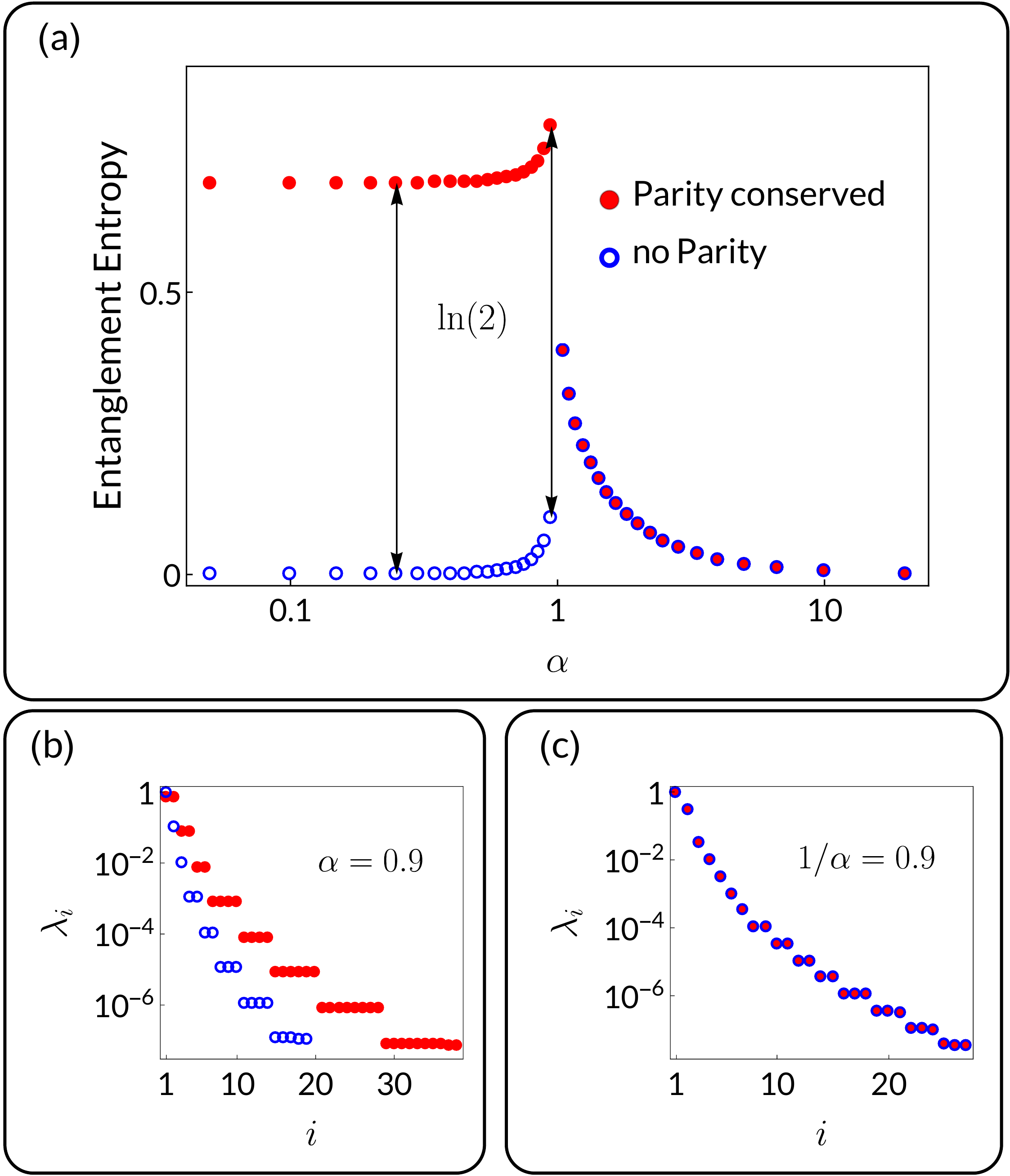}
    \caption{(Color Online) (a) Entanglement entropy of the parity-conserving and parity-non-conserving MPS as a function of the dimensionless transverse field, $\alpha$. The model has a quantum critical point at $\alpha=1$ where the entanglement entropy diverges logarithmically. For $\alpha>1$ the ground state has a definite parity and both ansatze agree with one another. For $\alpha<1$, the model is in the symmetry-broken phase. The parity-conserving ansatz must remain in an even-parity state, and thus converges to a GHZ state. The parity-non-conserving state is able to break the $\mathbb{Z}_2$ symmetry and converge to a lower-entropy state. As shown in the figure, the difference in entropy for $\alpha<1$ is precisely $\ln(2)$. (b) Singular values of both ansatze at $\alpha=0.9$. The values mirror one another, but the parity-conserving ansatz has exactly double the number of singular values. (c) Singular values of both ansatze at $1/\alpha=0.9$. Here both ansatze converge to the same definite-parity state, and hence their singular values are identical.}
    \label{fig:tfising}
\end{figure}

We use the infinite DMRG algorithm (iDMRG) to find the ground state of Eq.~(\ref{tfiham}) as a function of $\alpha$.    We separately run the algorithm with and without enforcing parity conservation, using appropriate initial conditions.
In Fig.~\ref{fig:tfising}(a) we plot the resulting entanglement entropy
across a bipartition of the infinite chain.
%, when the chain is bisected. 
%of the Ising model ground state 
%as a function of $\alpha$. 
Filled red dots denote the behavior of the parity-conserving MPS, while open blue circles show the non-parity-conserving results. For $\alpha>1$ the ansatze converge to the same state, which has zero entanglement entropy as $\alpha\to\infty$. This parent state is simply a product state with all spins oriented in the $\hat z$ direction.
%, and both data points agree because this phase has definite (even) parity. 
The entanglement entropy diverges at the critical point, $\alpha=1$. For $\alpha<1$, the two simulations converge to distinct, degenerate ground states. As shown by the arrows, the parity-conserving ansatz has exactly $\ln(2)$ more entanglement entropy at every point with $\alpha<1$. This relationship is expected when the parity conserved state is a simple superposition of the two symmetry-broken ground states. {%\color{blue} 
We note that this feature of unconstrained DMRG, in which the algorithm converges to the minimally-entangled degenerate ground state, is generic and has been recognized in the context of topological systems~\cite{jiang2012,jiang2013}.}

We investigate this correspondence more closely in Figs.~\ref{fig:tfising}(b) and (c), where we plot the spectrum of singular values, $\lambda_i$, at representative points in both phases: $\lambda_i^2$ are the eigenvalues of the reduced density matrix when one traces over half the chain.
Figure~\ref{fig:tfising}(b) shows the spectrum at a representative point in the symmetry broken  phase. The spectrum is effectively doubled by conserving parity -- each blue singular value matches up with exactly two red singular values in each degenerate plateau. This is precisely what one would expect by taking a superposition of symmetry-broken states. Note that the slight offset between corresponding blue and red plateaus is due to the normalization condition, $\sum_i\lambda_i^2=1$, and the fact that the parity-conserving ansatz has twice as many singular values. By contrast, Fig.~\ref{fig:tfising}(c) shows that the singular values of both states match up perfectly when $\alpha>1$.

A consequence of this spectral doubling is that
%Because of the spectral doubling, 
the symmetry-preserving MPS in the symmetry-broken phase needs twice the bond dimension to yield the same accuracy as the wavefunction which explicitly breaks the symmetry.  The MPS  tensors therefore contain four times as many matrix elements, only {\em half} of which are eliminated by the block-sparseness condition in Eq.~(\ref{eq:sparsity}).  Thus, instead of making  the calculation more efficient, enforcing parity conservation requires storing twice as many matrix elements.  On the paramagnetic side of the transition, the situation reverses, and the parity-conserving ansatz requires half as many elements.  

The transverse field Ising model is simple,  and the algorithmic costs/benefits here are small.  Nonetheless, it provides a clear illustration of how spontaneous symmetry breaking interacts with conservation laws in DMRG.

%Importantly, the number of singular values plotted in Figs.~\ref{fig:tfising}(b) and (c) is equal to the bond dimension of the MPS tensors, $\chi$. One will often keep all singular values above some threshold lower bound in order to determine system parameters with a desired precision. As tensor manipulations used in optimization routines generically scale polynomially in $\chi$, the bond dimension is directly related to both the algorithmic efficiency as well as the compactness of storage. The bond dimension of the parity-conserving ansatz is double that of the non-conserving ansatz in the symmetry-broken phase at any given precision, despite the fact that it contains no more information about the ground state (other than the global $\mathbb{Z}_2$ symmetry, which was known from the outset). In this way, the use of the parity-conserving ansatz is less efficient when describing the symmetry-broken phase of the transverse-field Ising model.

\subsection{\label{sec:BoseHubbard}Bose-Hubbard model}

The Bose-Hubbard model is a paradigmatic strongly-interacting model of lattice bosons. In one dimension (1D) the Hamiltonian is
\begin{equation}
    \mathcal{H}_{\rm BH}=-t\sum_{j}(a^\dagger_{j}a_{j+1}+h.c.)+U\sum_j n_{j}n_{j}
    \label{eq:Hbh}
\end{equation}
where $a^{(\dagger)}_j$ is a bosonic annihilation (creation) operator on site $j$ of a lattice, and $n_j=a^\dagger_ja_j$ is the number operator. We focus on the superfluid phase, which in 1D is a critical phase described by Tomonaga-Luttinger liquid theory~\cite{haldane1981,cazalillaReview,giamarchi,Kiely2022}. Unlike a Bose-Einstein condensate, it does not  spontaneously  break ${\rm U}(1)$ gauge invariance. There is, however, quasi-long range order corresponding to a power law decay of the single particle density matrix.  In contrast  to the phases in Sec.~\ref{sec:TFIsing}, it is not \textit{a priori} obvious whether this superfluid phase would be better described by a variational wavefunction that breaks or conserves particle-number conservation.

The most interesting part of the phase diagram is near the BKT transition at the tip of the Mott lobe.  Thus 
we focus on the point $U/t=3$ with an average of $\bar n=1$ particles per site. 
%This point is very close to the BKT transition at the tip of the Mott lobe, and hence requires a careful scaling analysis. 
We use the standard iDMRG algorithm.   
For our particle-number-conserving simulations, fixing the average density is trivial, while in our unrestricted simulations we add an extra step in each iteration which corrects the chemical potential, $\mu$.  This procedure is described in Appendix~\ref{sec:mu}.  
%In Sec.~\ref{sec:FermiHubbard} we discuss  some  of the issues  related to using particle number conservation at fractional filling.  
As described in Ref.~\cite{Kiely2022}, this gapless, critical phase is best analyzed using ``finite entanglement scaling," meaning that one understands the properties of the state by considering a sequence of bond dimensions, $\chi$.
%For unit filling, iDMRG simulations with particle number conservation can be carried out with a single-site unit cell -- we defer a discussion of fractional fillings to Sec.~\ref{sec:FermiHubbard}.

\begin{figure}
    \centering
    \includegraphics[width=3.375in]{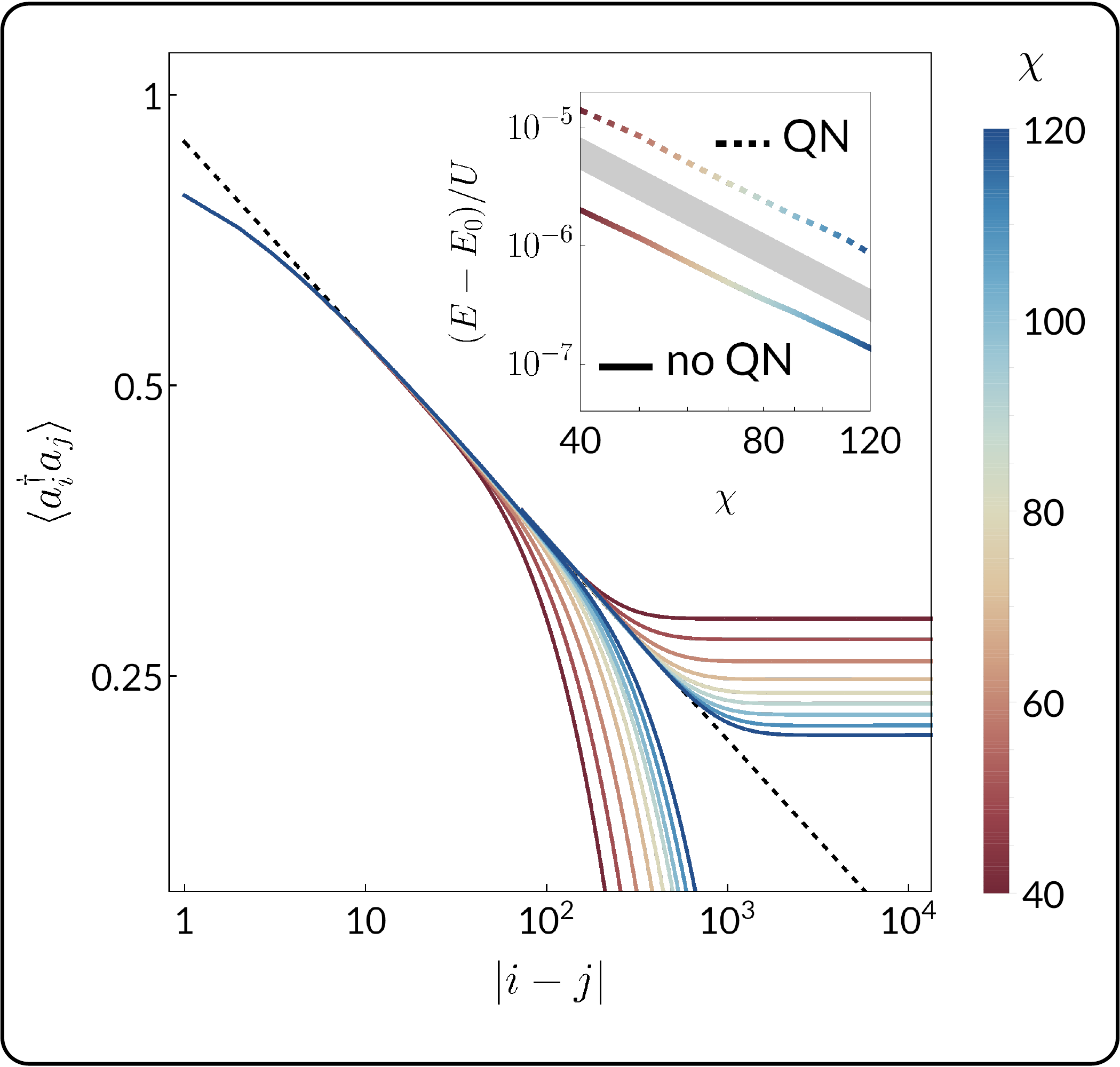}
    \caption{(color online) Density matrix of the 1D Bose-Hubbard model, $\langle a^\dagger_ia_j\rangle$, versus spatial separation, $|i-j|$, on a log-log scale for a variety of bond dimensions $\chi$, denoted by color. Model parameters are $U/t=3$, $\bar n=1$. Curves utilizing the block-sparse particle number conserving MPS bend downward, while the unrestricted dense MPS curves exhibit a plateau corresponding to Bose-Einstein condensation. Dashed line shows the asymptotic power-law decay based on a scaling analysis (see Appendix~\ref{sec:scaling}). 
    %Clearly the dense MPS curves capture the asymptotic behavior out to significantly longer distances than the number-conserving curves.
    Inset: Variational energy in units of $U$ versus bond dimension on a log-log scale for dense and sparse matrix product states. The dense MPS achieves a lower energy for all bond dimensions, while both curves exhibit scaling consistent with Eq.~(\ref{eq:varEn}), whose slope is given by the thick gray line.}
    \label{fig:boseHubbard}
\end{figure}

Our results are summarized in Fig.~\ref{fig:boseHubbard}. In the main panel we plot the density matrix $\langle a^\dagger_ia_j\rangle$ as a function of spatial separation $|i-j|$. For the number-conserving ansatz, the correlation function falls off exponentially at sufficiently long distances.  For the dense ansatz, the correlation function instead approaches a constant. This constant corresponds to a Bose-Einstein condensate, indicating that the finite-bond-dimension approximation spontaneously breaks the symmetry even though exact ground state is critical.  One typically refers to this phenomenon as ``quasicondensation," 
{%\color{blue} 
characterized by a quasi-condensate density, $\rho_{qc}$. We discuss this at greater length in Sec.~\ref{sec:Disc}.}
%which we discuss at greater length in Sec.~\ref{sec:Disc}.
The dashed black line shows the asymptotic power-law scaling of the density matrix,
\begin{equation}
    \langle a^\dagger_ia_{j}\rangle\propto|i-j|^{-K/2},
    \label{eq:dMatLL}
\end{equation}
where we used a scaling analysis to find the Luttinger parameter $K$  (see Appendix~\ref{sec:scaling}). The dense MPS better captures the correlations: while both curves eventually bend away from the dashed line, the dense MPS curves show approximate power-law decay out to distances almost an order of magnitude larger than those of the sparse MPS.
%As shown by the inset of Fig.~\ref{fig:boseHubbard}, the energy is also lower for the ansatz which does not conserve quantum numbers. 

The inset of Fig.~\ref{fig:boseHubbard} shows the variational energy of the dense and sparse ansatze as a function of bond dimension on a log-log scale. The dense MPS has a substantially lower energy at each bond dimension, while both curves exhibit power-law scaling of the form
\begin{equation}
    E(\chi)=E_0+A/\chi^{2\kappa}+\ldots\hspace{1cm}A>0,
    \label{eq:varEn}
\end{equation}
{%\color{blue} 
where $E_0$ is the true ground-state energy in the thermodynamic limit.}
As argued in Ref.~\cite{pollmann2009}, one expects that $\kappa=6/(c+\sqrt{12c})$ for an MPS approximation of a conformal critical point with central charge $c$. The low-energy description of the Bose-Hubbard model takes the form of a single-component Luttinger liquid, which is a conformal field theory with $c=1$. This theoretical prediction is given by the shaded gray line in the inset, clearly showing that the data is in close agreement with Eq.~(\ref{eq:varEn}).  The dense ansatz is roughly an order of magnitude more accurate for the same bond dimension.

%XXXXXXXXXX

%In other words, {\color{red} the correlation length}, which can be roughly understood as the distance beyond which the density matrix exhibits non-power-law features, is substantially larger for the dense MPS.

% The inset of Fig.~\ref{fig:boseHubbard} shows the variational energy of the dense and sparse ansatze as a function of bond dimension. The dense MPS has a substantially lower energy at each bond dimension. Schematically, the energy as a function of bond dimension is given by~\cite{pollmann2009}
% \begin{equation}
%     E(\chi)=E_0+A/\xi^2(\chi)+\ldots\hspace{1cm}A>0,
%     \label{eq:varEn}
% \end{equation}
% so the difference in energy can be largely understood as a consequence of the longer correlation length $\xi(\chi)$ of the dense MPS. As argued in Ref.~\cite{pollmann2009}, one expects $\xi(\chi)\propto \chi^\kappa$ where $\kappa=6/(c+\sqrt{12c})$ for an MPS approximation of a conformal critical point with central charge $c$. The Bose-Hubbard model is described by a single-component Luttinger liquid field theory and hence has $c=1$. We verify from $\partial E(\chi)/\partial \chi$ that the scaling analysis in Eq.~(\ref{eq:varEn}) is valid for the bond dimensions shown.

In contrast to the energy, the correlation length behaves counterintuitively.  We define 
\begin{equation}\label{xi}
\xi^2 =\frac{\sum_j j^2\left(\langle a_j^\dagger a_0\rangle -\langle a_j^\dagger\rangle\langle a_0\rangle\right)}{\sum_j \langle a_j^\dagger a_0\rangle -\langle a_j^\dagger\rangle\langle a_0\rangle},
\end{equation}
which is the characteristic length-scale of the fluctuations.  Despite the fact that the number non-conserving ansatz yields a density matrix which is closer to the exact result (which has an infinite correlation length), its correlation length is shorter.  This unexpected result is a consequence of subtracting off the constant term in Eq.~(\ref{xi}).

In addition to the ${\rm U}(1)$ symmetry described here, at the BKT point
%We note as well that 
the Bose-Hubbard model exhibits an emergent $\mathbb{Z}_2$ particle-hole symmetry. 
As the number-conserving ansatz encodes particle and hole fluctuations with different singular values, this implies that near the BKT point many of its singular values will have nearly-degenerate partners. Similar degeneracies have been used to detect forms of order~\cite{Li2008,thomale2010,pollmann2010}. They also indicate that the ansatz contains redundant information. These degeneracies do not show up in the singular values for the non-conserving ansatz.
%As the non-conserving ansatz does not have to use different singular values for number fluctuations, its entanglement spectrum has no such degeneracy. Analogous to the spectra presented in Fig.\ref{fig:tfising}(b), the lack of a $\mathbb{Z}_2$ redundancy implies that the dense ansatz is a more efficient variational ansatz.

%To conclude this section, we emphasize that the dense MPS does in fact encode particle-hole degeneracies in the singular values when evaluated at a point inside the Mott lobe. In fact, we find that the dense ansatz converges to the exact same state as the number-conserving ansatz (within numerical precision) in the Mott phase. This indicates that the absolute variational minimum for fixed $\chi$ is contained within the subspace of sparse MPS within the Mott insulating phase. By contrast, the absolute variational minimum in the superfluid phase is a quasicondensing dense MPS (for finite $\chi$), and hence the sparse variational minimum must be interpreted as a local minimum.

\subsection{\label{sec:FermiHubbard}Fermi-Hubbard model}

The 1D Fermi-Hubbard model describes spin-1/2 lattice fermions  with on-site interactions and Hamiltonian
\begin{equation}
    \mathcal{H}_{\rm FH}=-t\sum_{j,\sigma}(c^\dagger_{j,\sigma}c_{j+1,\sigma}+h.c.)+U\sum_j n_{j,\uparrow}n_{j,\downarrow}.
    \label{eq:Hfh}
\end{equation}
Here $c^{(\dagger)}_{j,\sigma}$ is a fermionic annihilation (creation) operator for a particle with spin $\sigma$ on site $j$ and $n_{j,\sigma}=c^\dagger_{j,\sigma}c_{j,\sigma}$ is the number operator. Like the 1D Bose-Hubbard model, the ground state of the 1D Fermi-Hubbard model is either a Mott insulator or a Luttinger liquid. We will again focus on the latter phase.
%, as it is well-established that number conservation is a useful constraint in the former. 
This model is exactly solvable via the Bethe ansatz~\cite{lieb1968,schultz1990}.

At half filling (one particle per site) this model is in the Mott insulator phase for any $U/t>0$.  Thus we work at quarter filling and zero net magnetization,
%The 1D Fermi-Hubbard model with only nearest-neighbor tunneling has a nesting-driven instability at half filling such that the ground state is a Mott insulator for any infinitesimal $U/t>0$~\cite{lieb1968}. Working at fractional fillings guarantees that the system is in the Luttinger liquid phase. We perform iDMRG simulations to find the variational ground state for $U/t=4$ at quarter-filling and zero net magnetization: 
$\bar n_\uparrow = \bar n_\downarrow=1/4$. Our block-sparse simulations conserve both the total particle number and the total magnetization. Number-conserving simulations at a fractional filling $p/q$, where $p$ and $q$ are integers, requires a unit cell of length $qm$ sites where $m\in\mathbb{Z}^+$~\footnote{It is in fact possible to work with smaller unit cell sizes, but this will effectively encode the larger unit cell by cycling through different blocks of the MPS. This enlarges the bond dimension artificially without improving accuracy, which is detrimental to algorithmic efficiency.}. 
The dense MPS simulations have no restriction on the allowed unit cell size.
For the purpose of providing a reliable comparison between methods, we perform both the number-conserving and non-number-conserving simulations with a unit cell of 4 sites. A good discussion of multi-site iDMRG can be found in Ref.~\cite{mcculloch2008}.
%Note, however, that the dense MPS simulation can be performed for a variety of unit cells.

\begin{figure}
    \centering
    \includegraphics[width=3.375in]{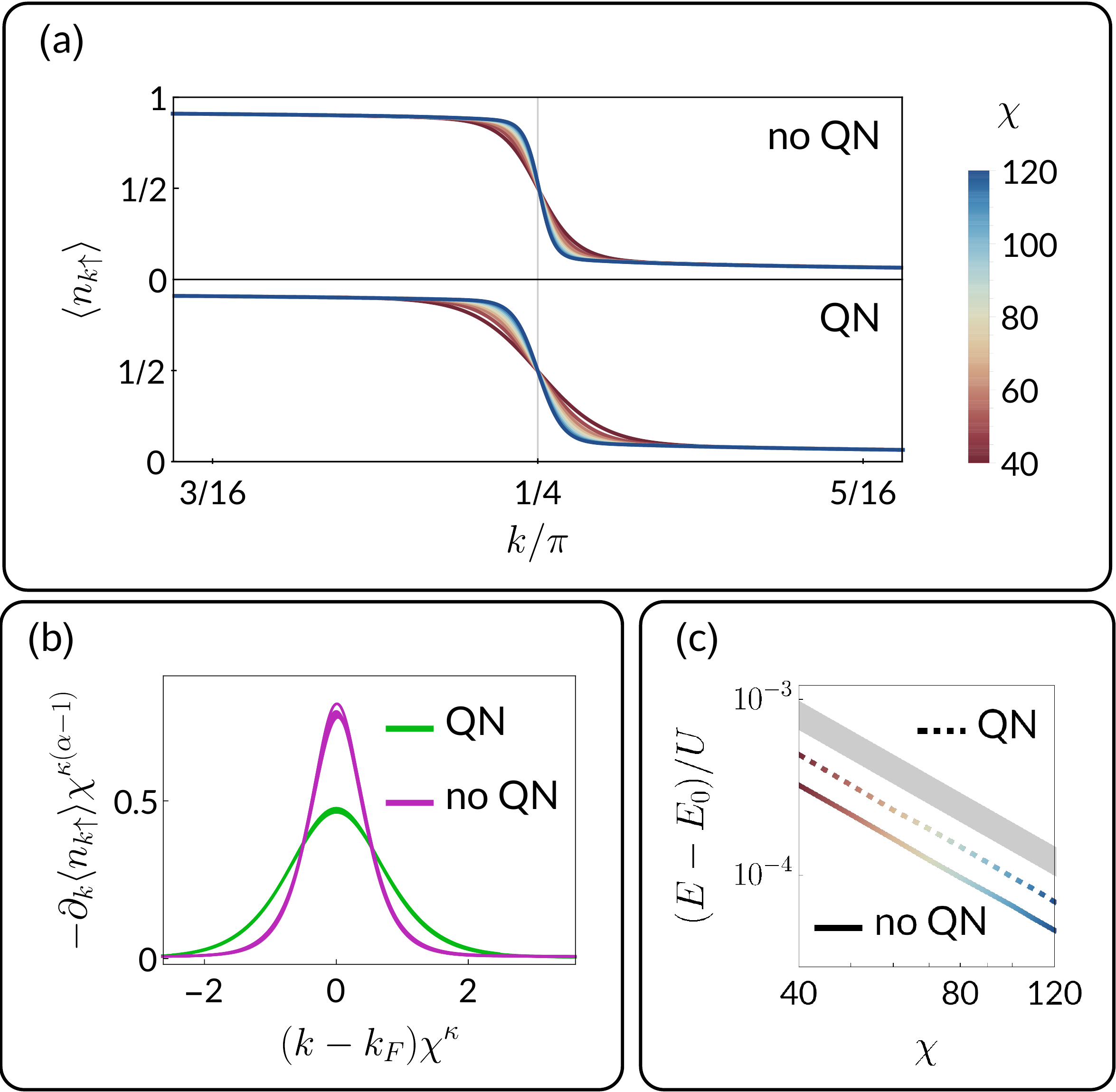}
    \caption{(color online) (a) Momentum distribution function for $\uparrow$ spins, $\langle c^\dagger_{k\uparrow}c_{k\uparrow}\rangle$, in the vicinity of $k_F=\pi/4$ for the 1D Fermi Hubbard model. Model parameters are $U/t=4$, $\bar n_\uparrow=\bar n_\downarrow=1/4$. Top and bottom plots show data from the dense and sparse ansatz. All curves display the expected power-law scaling (see Eq.~(\ref{eq:kfLL})) up to broadening of the power-law singularity due to the finite MPS correlation length. The distribution function is considerably sharper for the dense MPS, indicating a longer correlation length. (b) Derivative of the momentum distribution function, $\partial \langle n_{k\uparrow}\rangle/\partial k$, versus $k$. The axes have been rescaled by $\chi^{\kappa(\alpha-1)}$ and $\chi^\kappa$, respectively, where $\kappa$ and $\alpha$ are defined in the main text. The dense and sparse curves from panel (a) exhibit distinct scaling collapses, as shown. (c) Variational energy in units of $U$ versus bond dimension on a log-log scale for dense and sparse matrix product states. As with the Bose-Hubbard model, we see the dense MPS provides a lower variational energy and that the energies of both ansatze are described by a power law consistent with Eq.~(\ref{eq:varEn}) (gray line).}
    \label{fig:fermiHubbard}
\end{figure}

Figure~\ref{fig:fermiHubbard}(a) shows the momentum distribution function for up spins, $\langle n_{k\uparrow}\rangle$, in the vicinity of $k_F=\pi/4$. Both data sets show a step at $k_F$ that grows increasingly sharp with bond dimension. Note that on this scale the distribution never goes to $0$ or $1$.  The step height is somewhat analogous to the Fermi liquid quasiparticle weight $Z$.
As $\chi\to\infty$ the distribution function near $k=k_F$ should approach~\cite{giamarchi}
%Both data sets qualitatively reproduce the sharp singularity at $k_F$ that is emblematic of fermionic Luttinger liquids,
\begin{equation}
    \langle n_{k,\sigma}\rangle\approx\frac{1}{2}-{\rm sign}(k-k_F)|k-k_F|^\alpha,
    \label{eq:kfLL}
\end{equation}
%up to corrections in the vicinity of $k_F$.
%Deviations from this precise form are due to the finite correlation length of the MPS at fixed bond dimension, which we will discuss shortly. 
where the exponent in the power-law singularity depends on the Luttinger parameter for charge degrees of freedom, $K_\rho$:
\begin{equation}
    \alpha=(K_\rho+1/K_\rho-2)/4.
\end{equation}
For $U/t=4$ and $\bar n_\uparrow=\bar n_\downarrow=1/4$ the Bethe ansatz solution gives $K_\rho\approx 1.4$~\cite{schultz1990}.

At $k=k_F$, the derivative $\frac{\partial \langle n_{k\uparrow}\rangle}{\partial k}$ in Eq.~(\ref{eq:kfLL}) diverges.
For finite $\chi$, this singularity is cut off and one instead expects $\frac{\partial \langle n_{k\uparrow}\rangle}{\partial k}\sim \chi^{\kappa (1-\alpha)}$ where 
$\kappa=3/(1+\sqrt{6})$ is the conformal scaling exponent corresponding to a central charge $c=2$~\cite{pollmann2009}.  The width of the deviation from Eq.~(\ref{eq:kfLL}) scales as $\delta q \propto \chi^{-\kappa}$.  
Figure~\ref{fig:fermiHubbard}(b) demonstrates the resulting scaling collapse:  For a given ansatz, all of the curves from panel (a) lie on on top of one-another.  We use the theoretical values of $\kappa$ and $K_\rho$, without any free parameters.

Strikingly, in Fig.~\ref{fig:fermiHubbard}(b), the dense and sparse MPS exhibit two distinct scaling collapses, the former notably sharper than the latter. Thus, while both data sets exhibit the expected conformal scaling, the dense MPS yields wavefunctions with a sharper singularity at $k_F$.
%, which is indicative of a larger correlation length. 
%the expected conformal scaling is achieved for both data sets even at these low bond dimensions, but the dense MPS provides a superior variational wavefunction at any given bond dimension. 
These two scaling collapses can be made to line up with one another by rescaling the bond dimension
by a factor of 1.8.  

%s of either the dense or sparse data sets by a constant factor. Such a rescaling allows us roughly define an equivalent bond dimension, i.e. the correlation function produced by a dense MPS with bond dimension $\chi$ is equivalent to that produced by a sparse MPS with bond dimension $1.8\chi$.

In Fig.~\ref{fig:fermiHubbard}(c) we plot the variational energy as a function of the bond dimension on a log-log scale. We again find that the energy obtained by the dense MPS is  lower than that of the sparse MPS.
%(note the increase in scale compared to Fig.~\ref{fig:boseHubbard}).  
Fixing the bond dimension, the ratio of the errors in the energy for the two ansatze is roughly 1.5.  The shaded gray line denotes the scaling behavior in Eq.~(\ref{eq:varEn}), which is clearly consistent with both data sets.

%although 
%The difference in their energies  is a small fractio
%both variational energies are significantly larger that the extrapolated ground state energy. In this sense, the relative improvement that comes from using a dense MPS is less significant when compared to the Bose-Hubbard example. The shaded gray line denotes the scaling behavior in Eq.~(\ref{eq:varEn}), which is clearly consistent with both data sets.

%(note that the scale on the energy axis is $\sim 10\times$ that of the inset in Fig.~\ref{fig:boseHubbard}). As with the Bose-Hubbard model, we compute $\partial E(\chi)/\partial \chi$ and verify that deviations from the asymptotic ground state energy are of the form $\delta E\propto \xi^{-2}\propto \chi^{-2\kappa}$, in accord. This provides further evidence that the advantage of a dense MPS can be attributed to a longer correlation length at a fixed bond dimension.

\section{\label{sec:Disc}Discussion}

In the symmetry broken phase of the transverse field Ising model, the two-fold-degenerate ground state manifold is spanned by symmetry broken states $|\pm\rangle$, or symmetry preserving states $|+\rangle \pm |-\rangle$.  The symmetry broken states have a smaller entanglement entropy, and hence can be described by a MPS with smaller bond dimension.  The matrices in the symmetry preserving MPS, however, are sparse.

The situation is more complicated in Secs.~\ref{sec:BoseHubbard} and~\ref{sec:FermiHubbard}, where we explored critical Luttinger liquid states.  In the thermodynamic limit these have infinite entanglement entropy, and hence an exact representation would require an MPS with infinite bond dimension.  For finite bond dimension, the dense MPS ansatz breaks the ${\rm U}(1)$ gauge symmetry and exhibits quasicondensation.
%displays a quasi-condensate.  
Similar to the Ising  model example, one can construct a number conserving state with density $\bar n$ by averaging over all values of the broken symmetry: $|\bar n\rangle = \int \!d\theta\,\exp(i\theta(\hat n-\bar n)) |\psi_0\rangle$.  Unfortunately, the $|\bar n\rangle$ constructed in this manner will have infinite bond dimension.  This points towards a more complex relationship between the number-conserving and symmetry-broken MPS approximants.  Nonetheless,  for a  fixed bond dimension, the  symmetry-broken wavefunction yields a more accurate energy.  As with the case of the transverse field Ising model, this increase in accuracy comes with the cost of requiring the use of dense matrices.

The symmetry breaking found at finite $\chi$ is analogous to the quasi-condensation seen in 1D Bose gases confined in traps of length $L$~\cite{Gangardt2003}.   Matrix product states with finite bond dimension always have a finite correlation length, $\xi$, and this length scale plays a similar role to $L$.
Just as our quasicondensate density vanishes as $\chi\to\infty$, these physical systems have $\rho_{qc}$ vanish as $L\to\infty$.

In the Fermi-Hubbard model, the quasicondensation discussed above corresponds to fictitious bosons which are constructed via a Jordan-Wigner transformation.  This therefore corresponds to a topological order in the fermionic system, which is revealed via a string correlation function.  There is no obvious way to experimentally measure this topological quasi-order.

Comparing the inset of Fig.~\ref{fig:boseHubbard} with Fig.~\ref{fig:fermiHubbard}(c), it is clear that the advantage gained from breaking the symmetry is larger for the bosons than for the fermions. As noted in Eq.~(\ref{eq:varEn}), 
the leading deviation of the variational energy is $\delta E=A \chi^{-2\kappa}$. Here $A$ is smaller for the dense ansatz, and the improvement in accuracy from using the dense ansatz is quantified by the dimensionless ratio $\Gamma=A_{\rm QN}/A_{\rm dense}$.  To achieve a fixed error in the energy, the number-conserving ansatz requires a bond dimension which is  $\Gamma^{1/2\kappa}$ times larger than the dense ansatz.  In our bosonic example (Sec.~\ref{sec:BoseHubbard}), $\Gamma=6.7$, while in the fermionic one (Sec.~\ref{sec:FermiHubbard}), $\Gamma=1.5$.

The reason for this difference is that the fermionic system has much smaller density fluctuations. In the number conserving ansatz, there is a configurational entropy associated with number fluctuations between two halves of the system,  requiring a larger bond dimension.  The scale of these number fluctuations is set by the Luttinger parameter:  A region of size $L$ will have fluctuations $\langle (\hat N-\langle\hat N\rangle)^2\rangle \sim K^{-1} \ln L$~\cite{song2010}.  In a MPS of fixed bond dimension, the correlation length $\xi$ plays the role of $L$.  In our examples $K^{\rm Bose}=0.42$ is much smaller than $K^{\rm Fermi}_\rho=1.4$.

\begin{figure}
    \centering
    \includegraphics[width=3.375in]{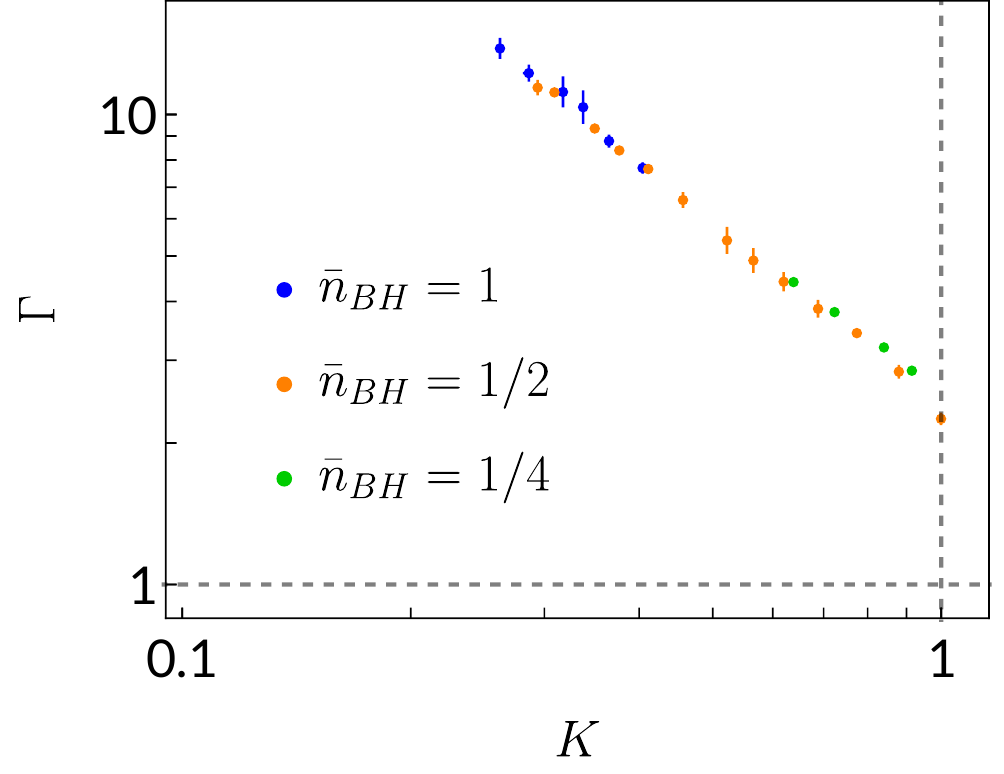}
    \caption{(color online) Ratio of leading coefficients in the energy scaling function (Eq.~\ref{eq:varEn}), $\Gamma=A_{\rm QN}/A_{\rm dense}$, versus the Luttinger parameter. Blue, orange and green data correspond to the Bose-Hubbard model at particle densities $\bar n = 1$, $1/2$, and $1/4$, respectively. Strikingly, these data lie on a single scaling function, independent of the microscopic parameters. The function exhibits a power-law divergence as $K\to 0$, where the ground state is proximate to a Bose-Einstein condensate, while it should approach $1$ as $K\to\infty$. }
    \label{fig:scaling}
\end{figure}

In Figure~\ref{fig:scaling} we show how $\Gamma$ depends on  Luttinger parameter for the Bose-Hubbard model at three different fillings: $\bar n=1$, $1/2$, and $1/4$. We find that these data collapse onto a single universal curve which diverges as a power law, $\Gamma\propto K^{-1.27(2)}$, for small $K$. The maximum value of $K$ in the superfluid phase of the Bose-Hubbard model is $K=1$, beyond which the system undergoes a Mott transition. If one were to continue the scaling function out to $K>1$, e.g. with the inclusion of long-range interactions, one would expect the power-law behavior to break down so that $A_{\rm dense}/A_{\rm QN}\to 1$ as $K\to\infty$.

The metallic phase of the 1D Fermi-Hubbard model with $U>0$ has distinct Luttinger parameters for spin ($K_\sigma$) and charge ($K_\rho$) degrees of freedom. The spin Luttinger parameter is fixed at $K_\sigma=1$, while the charge Luttinger parameter $K_\rho>1$. Both spin and density fluctuations are relevant here, so the fermionic results do not collapse onto the bosonic data in Fig.~\ref{fig:scaling}.

\section{\label{sec:conclusion}Conclusions}

Conservation laws allow one to write matrix product states in a block-sparse manner (Eq.~(\ref{eq:sparsity})). It is not, however, always favorable to take advantage of this structure. {%\color{blue} 
For example}, if the ground state spontaneously breaks the symmetry then the resulting MPS contains redundant information whose only purpose is to impose the constraint.

{%\color{blue} 
These considerations are particularly interesting for
critical Luttinger liquid phases, where the symmetry is {\em almost} broken.  We find that for fixed bond dimension one more accurately estimates the ground-state energy by using a dense ansatz that does not rely on the symmetry.
The benefits of the dense ansatz
%of using a symmetry preserving MPS ansatz 
are greatest when the Luttinger parameter, $K$, is small.    
%The advantage is greatest for small $K$.  This feature is seen in both the Bose and Fermi Hubbard models.

Although more accurate at a fixed bond dimension, the dense ansatz  requires more computational resources.   
For high accuracy calculations ($\delta E/U<10^{-6}$) the sparse ansatz runs faster and uses less memory.  At moderate accuracy, however, the dense ansatz is more efficient.  The threshold value of $\delta E/U$ falls with decreasing $K$.

%are generically more complicated, as the scaling properties of the ansatz with bond dimension must be considered. Nonetheless, we have shown that these states can be modeled considerably more accurately (at fixed bond dimension) with the use of dense tensors. The degree to which this is true depends on the Luttinger parameter. As we discuss in Appendix~\ref{sec:efficiency}, this can translate into practical gains in efficiency in certain parameter regimes.

%We emphasize that these 
Our results %for critical systems 
are relevant for a wide variety of systems. 
Any gapless system 
%As the MPS ansatz is fundamentally one-dimensional, use of this variational wavefunction to model gapless systems 
(including quasi-2D geometries) will invariably have critical Luttinger-liquid-like features 
when modelled using a MPS.}
%in the long-wavelength limit.}
Moreover, our considerations apply to all tensor network approaches~\cite{bauer2011,singh2011}. Efficient numerical calculations require an awareness of the interplay between spontaneous symmetry breaking and conservation laws.

\section{Acknowledgements}
This work was supported by the NSF Grant No. PHY- 2110250.

\appendix

{%\color{blue}
\section{\label{sec:efficiency}Implementation dependent metrics}
Here we compare {\em computer memory usage} and {\em wall time per iteration}
for iDMRG calculations of the ground state of the Bose Hubbard model, using either a sparse or dense representation of the tensors in the matrix product state.  These metrics depend on hardware and implementation details.  Here we use the 
iDMRG algorithm~\cite{SCHOLLWOCKreview,mcculloch2008} using the ITensor C++ library~\cite{itensor} compiled with Intel MKL on a single core without multithreading.  The code for these calculations involves only minor tweaks to the native iDMRG code on ITensor~\cite{stoudenmireIDMRG}.  Despite the implementation dependence, we expect qualitative features to be generic.

The main conclusions are: 
(1) For small Luttinger parameter, $K$, it is favorable to use the dense MPS ansatz, unless one targets an extremely high accuracy.  For moderate  accuracies, the dense ansatz takes less memory and results in a faster calculation.  This behavior is analogous to the transverse field Ising model in the symmetry broken phase.  (2) For large $K$ it is always favorable to use the sparse ansatz.  This behavior is analogous to the transverse field Ising model in the paramagnetic phase.

%As discussed in the main text, there are a variety of non-overlapping ways to determine whether a numerical technique is appropriate for any given problem. In this section, we provide a brief discussion of two alternative ways to compare number-conserving and non-conserving MPS ansatze: {\em computer memory usage} and {\em wall time per iteration}. We emphasize that, while both of these metrics are concrete and practical constraints for any given researcher, they are sensitive to one's hardware as well as the libraries used for tensor storage and contractions. All the calculations in this paper are performed with the iDMRG algorithm~\cite{SCHOLLWOCKreview,mcculloch2008} using the ITensor C++ library~\cite{itensor} compiled with Intel MKL. While prior calculations have utilized multithreading, the wall time calculations in this section were performed on a single core.
%The code for these calculations involves only minor tweaks to the native iDMRG code on ITensor~\cite{stoudenmireIDMRG}.

\subsection{Memory Usage}

Each tensor in our dense MPS ansatz requires storing $N_{\rm tot}=\chi^2 d$ numbers.  Here $d$ is the dimension of the local Hilbert space:  In the Bose Hubbard model, $d=n_{\rm max}+1$ is set by the maximum number of particles that we allow on a site.  For a fixed $\chi$, the sparse representation requires a smaller $N_{\rm tot}$, as we do not need to store the entries which vanish due to symmetry.  To compare the memory usage of the two approaches, we define $\chi_{\rm eff}=\sqrt{ N_{\rm tot}/d}$.  For our Bose Hubbard calculations we find that for moderate $\chi\lesssim 400$ there is a nearly linear relationship between $\chi$ and $\chi_{\rm eff}$.

In Fig.~\ref{fig:chiEff} we show the 
accuracy of the dense iDMRG energy as a function $\chi$ (solid curves) and the sparse iDMRG energy as a function of $\chi_{\rm eff}$ (dotted curves).  For $K=0.42$ the sparse ansatz requires a smaller memory footprint to achieve the same accuracy (the red dotted curve lies below the solid red curve).  Conversely, for $K=0.11$, the sparse ansatz requires a larger footprint.  These observations are in line with the arguments from Sec.~\ref{sec:ConsLaws}, which suggest that at small $K$, where we are proximate to a Bose-Einstein condensate, the dense ansatz can more efficiently encode the quantum state.  For larger $K$ the advantage goes to the sparse ansatz.

Figure~\ref{fig:chiEff} also plots the sparse data versus $\chi$ as dot-dashed lines which always lie above the solid lines:  
For a fixed $\chi$, the dense ansatz has more degrees of freedom, and hence yields a lower variational energy.   As expected, the advantage is greatest for small $K$.

%From a practical standpoint, a quantum number-conserving MPS is useful because its block-sparse structure allows one to divide its storage into smaller components. To illustrate the benefit of this, let's compare a dense MPS with bond dimension $\chi$ and physical dimension $d$ with a number-conserving MPS with the same total linear dimensions. For simplicity, we'll assume that the number-conserving MPS is composed of $N$ blocks, each of the same size. A computer must store the dense MPS tensor as a set of $\chi^2d$ numbers, as each can in principle be non-zero. The sparse MPS, however, can be stored as a set of $N$ independent tensors, and hence only has a memory cost that scales as $\chi^2d/N$. In general, the block structure is non-trivial and can depend on a variety of parameters. We quantify the memory usage of the number-conserving MPS by summing over blocks to find the total number of free parameters in a given tensor, $N_{\rm tot}$, and then define an effective bond dimension $\chi_{\rm eff}=\sqrt{N_{\rm tot}/d}$.

\begin{figure}
    \centering
    \includegraphics[width=3.375in]{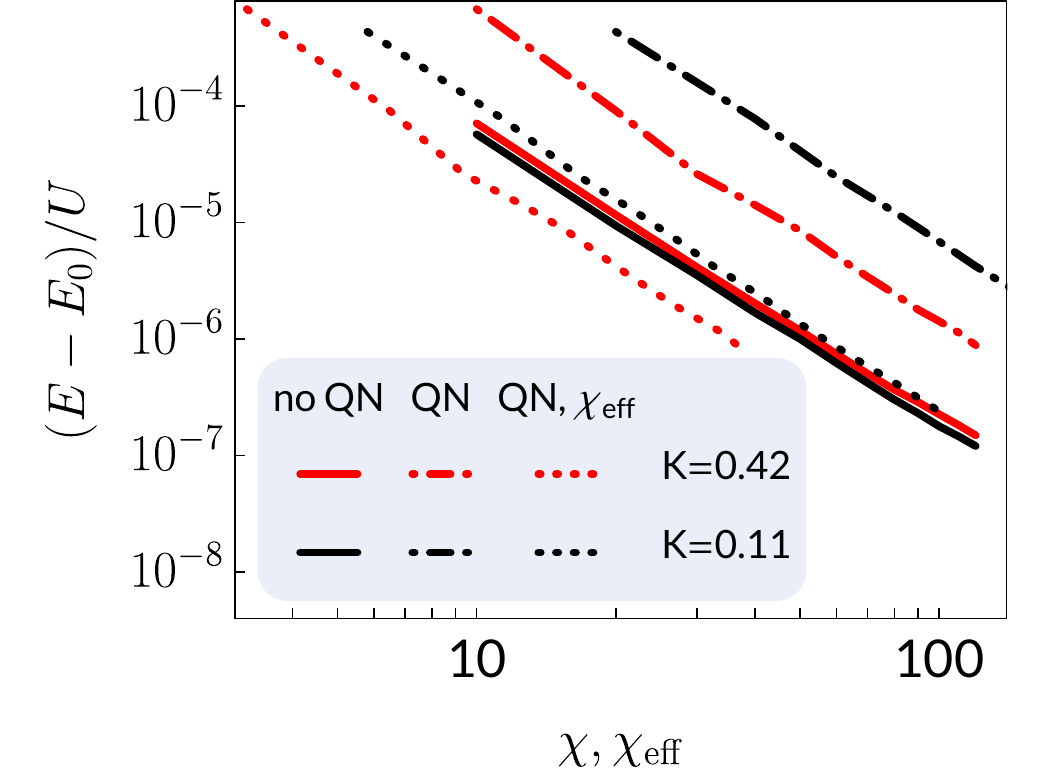}
    \caption{(color online) Variational energy of the 1D Bose-Hubbard model ($E$), measured relative to the extrapolated ground state energy ($E_0$), as a function of bond dimension $\chi$. 
    Line color corresponds to different values of $t/U$, resulting in Luttinger parameters $K=0.42,0.11$, shown as red or black (c.f. inset of Fig.~\ref{fig:boseHubbard}).
    Solid lines: dense ansatz, with $N_{\rm tot}=\chi^2 d$ degrees of freedom, where $d$ is the size of the local Hilbert space ($d=8$ for $K=0.11$ and $d=6$ for $K=0.42$); Dot-dashed lines: Sparse number conserving ansatz, with fewer degrees of freedom $N_{\rm tot}$.  For a fixed $\chi$ the dense ansatz always yields a smaller error, but the benefit decreases with increasing $K$.
    Dotted lines: Sparse number conserving ansatz, plotted vs. effective bond dimension $\chi_{\rm eff}\equiv \sqrt{N_{\rm tot}/d}$, which is a measure of the memory required to store the state.  For $K=0.42$ and a fixed memory footprint, the sparse ansatz yields more accurate results.  For $K=0.11$ there is little difference, but the dense ansatz is slightly more accurate.
    %Data is shown for the number-conserving and number-non-conserving ansatze for two different points on the phase diagram, characterized by Luttinger parameters $K=0.42$ (c.f. inset of Fig.~\ref{fig:boseHubbard}) and $K=0.11$. Also shown is the variational energy of the number-conserving ansatz as a function of its effective bond dimensions, $\chi_{\rm eff}$, as defined in Appendix~\ref{sec:efficiency}. We see that, for $K=0.42$, the dense MPS achieves a better variational energy for a given total bond dimension but the sparse MPS has a lower energy for a given memory allocation. For $K=0.11$, however, the dense MPS achieves a better variational energy for a given total bond dimension {\em and} for a given memory allocation.
    }
    \label{fig:chiEff}
\end{figure}

\subsection{Wall time}

%Another way of comparing ansatze is to consider the total wall time. While this is often one's primary limitation in carrying out simulations, this metric is also the most implementation-specific: total wall time for a calculation can depend on the library used for tensor manipulations, the speed of one's processor, metrics used to define convergence, and the initial state one chooses, among myriad other factors. For the purpose of this section, we eliminate the latter two variables by considering just the total wall time for a single iteration of iDMRG at the conclusion of the algorithm (i.e. near convergence). This metric, wall time per iteration, is sufficient to draw a sharp distinction between number-conserving and non-conserving MPS ansatze.

%To gain some intuition, we again consider the sparse MPS of bond dimension $\chi$, composed of $N$ blocks of equal size. The most expensive tensor contraction (and hence the wall time) in the unconstrained iDMRG algorithm scales as ${\rm O}(\chi^3dw)$ where $w$ is the dimension of the Hamiltonian MPO ($w=4$ for the 1D Bose-Hubbard model). The use of block-sparse constraints can reduce this substantially -- in our cartoon example, the wall time would instead scale as ${\rm O}(\chi^3dw/N^2)$. 

\begin{figure}
    \centering
    \includegraphics[width=3.375in]{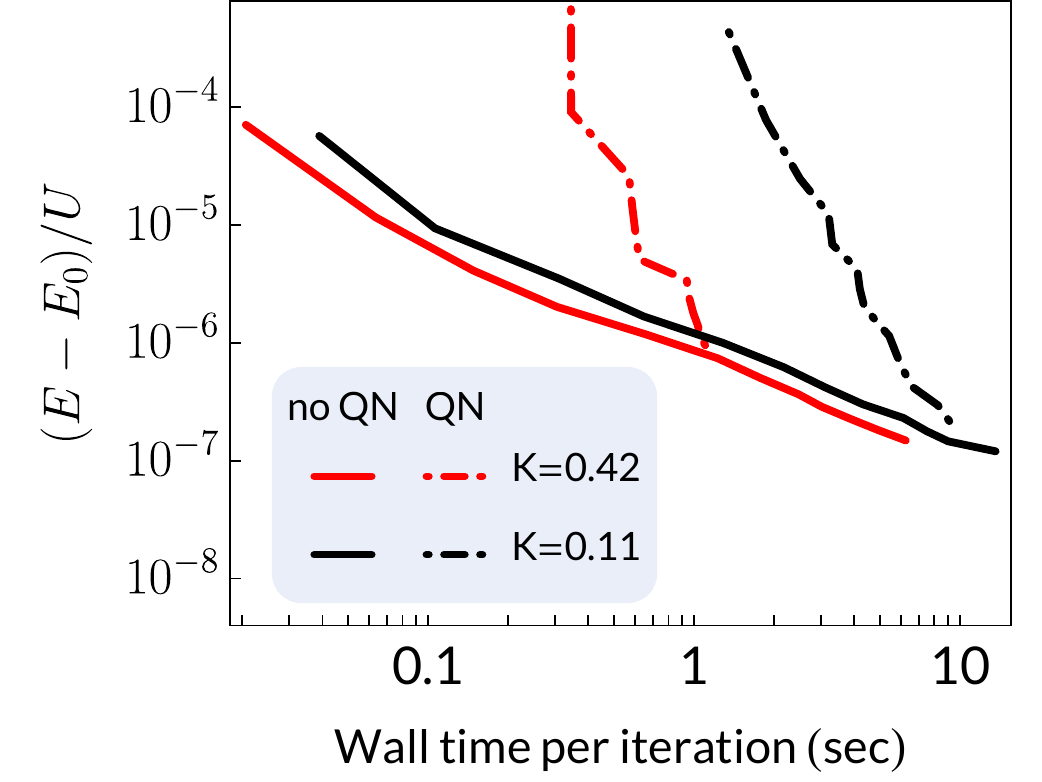}
    \caption{(color online) Variational energy of the 1D Bose-Hubbard model versus wall time per iteration, as bond dimension is varied. 
    %Solid lines: dense ansatz, Dot-dashed lines: sparse number conserving ansatz.  Color corresponds to the luttinger parameter:  Rede and number-non-conserving ansatze at $K=0.42$ and $K=0.11$. 
    Small bond dimensions correspond to short wall times, but low accuracy (see Fig.~\ref{fig:chiEff}).  The number conserving calculations (dot-dashed lines) have a steeper slope than those using dense matrices (solid lines), corresponding to a better scaling with bond dimension.  Nonetheless, for moderate accuracy calculations $(E-E_0)/U\gtrsim 10^{-6}$, the dense ansatz results in faster calculations.  The crossover point depends on $K$, and on implementation details.  Smaller $K$ benefits the dense calculation.       
%    For small bond dimensions, the number-conserving ansatz slows down because tensor contractions involve summing over different quantum number sectors. At large bond dimensions, sparse tensor contractions become much more efficient. The point at which these curves cross, beyond which the conserving ansatz is faster at achieving a given accuracy, shifts rightward (to lower energies and larger wall times) as $K\to 0$.
    }
    \label{fig:walltime}
\end{figure}

In Fig.~\ref{fig:walltime} we plot the same variational energies shown in Fig.~\ref{fig:chiEff}, but now versus the wall time per iteration. As previously noted, these times are highly dependent on implementation, and the actual number is not particularly relevant.  Nonetheless we can use these timings  to compare the performance of the two ansatze.

For the dense ansatz with $K=0.11$, one achieves an accuracy of $(E-E_0)/U=10^{-5}$ by taking $\chi=19$.  For these parameters, each iteration of the iDMRG algorithm takes a time of 0.1~sec.  By contrast, achieving the same accuracy with the sparse ansatz requires a larger $\chi=87$, and each iteratation takes significantly longer, 3.2~sec.  The sparse algorithm, however, scales much better with bond dimension, and if one has a target accuracy of $(E-E_0)/U=10^{-7}$, the two calculations take the same time.  If higher accuracy is required, the sparse ansatz is faster.

These features are illustrated in Fig.~\ref{fig:walltime} by the steeper slope of the sparse data.  As one moves to larger $K$, the relative performance of the sparse ansatz improves.  In particular,  at $K=0.42$, it is more time-efficient to use the sparse ansatz if one requres an accuracy smaller than $(E-E_0)/U=10^{-6}$.  Increasing $K$ farther continues to move this crossover point to lower accuracy.

}

\section{\label{sec:details}iDMRG Details}
In our calculations we
start with a product state, then implement the iDMRG algorithm with two-site updates to find MPS approximations of the ground state in the thermodynamic limit. Simulations are carried out using the ITensor library~\cite{itensor}.  Here we describe several technical details.

\subsection{Truncation Error}
For calculations of the transverse-field Ising model in Sec.~\ref{sec:TFIsing}, we increase the bond dimension as necessary until the properties of the state have all converged. We find that it is sufficient to reduce the truncation error $e_{\rm trc}\leq 10^{-12}$ to achieve convergence. One can interpret the data in Fig.~\ref{fig:tfising} as numerically exact results.
%For this reason, data points shown in Fig.~\ref{fig:tfising} correspond to matrix product states with different bond dimensions. 

As for the Bose-Hubbard (Sec.~\ref{sec:BoseHubbard}) and Fermi-Hubbard (Sec.~\ref{sec:FermiHubbard}) simulations, ``convergence" is no longer a meaningful criterion. The ground states are gapless critical states
and the long distance properties of the correlation functions cannot be
%that cannot be even qualitatively 
modeled by matrix product states with fixed bond dimension. As has been argued elsewhere~\cite{pollmann2009,Kiely2022}, however, features of the asymptotic ground state can be inferred by studying the behavior of variational wavefunctions as a function of the bond dimension. This is known as finite-entanglement scaling. For the data shown in Figs.~\ref{fig:boseHubbard} and~\ref{fig:fermiHubbard}, we increase the bond dimension from $\chi=40$ to $120$ in steps of $\Delta\chi=10$. We find that the truncation error scales as a power law of the bond dimension, $e_{\rm trc}\propto \chi^{-2\kappa}$, consistent with the Luttinger liquid scaling observed in other observables~\cite{pollmann2009}.

\subsection{Fixing the chemical potential}\label{sec:mu}
Throughout this paper, we calculate properties at fixed density.  This constraint is simple to incorporate into the sparse number conserving MPS ansatz.  For the dense ansatz, one instead has to specify a suitable
%While number-conserving iDMRG works at a fixed particle density, $\bar n$, the dense ansatz requires that one fix the density with a 
chemical potential, $\mu$, to fix the density at the desired value.
%therefore must estimate the point in parameter space at which the model has a given density in order to compare the two techniques.

To achieve our target density, $\bar n^*$, we vary the chemical potential in the early iterations of the dense iDMRG algorithm. 
The update procedure involves approximating the inverse compressibility based on measurements made in subsequent iterations:
\begin{equation}
    \frac{\partial \mu_i}{\partial \bar n_i}\approx\frac{\mu_i-\mu_{i-1}}{\bar n_i-\bar n_{i-1}}
\end{equation}
We then define the chemical potential for iteration $i+1$ based on the compressibility computed in iteration $i$:
\begin{equation}
    \mu_{i+1}=\mu_i+\alpha_i (\bar n^*-\bar n_i)\frac{\partial \mu_i}{\partial \bar n_i}.
\end{equation}
Here the  convergence factor, $\alpha$, controls how large of an update we allow from iteration to iteration.  Smaller $\alpha$ results in a more stable algorithm, but slower convergence.
In all our calculations we 
take $\alpha_i=0.1$. 
%In general, however, a more sophisticated routine would vary $\alpha_i$ based on the inverse Hessian computed in a given iteration.

\section{\label{sec:scaling}Scaling Analysis}
Here we describe how we use a scaling analysis of the single particle density matrix to extract the Luttinger parameter for the Bose-Hubbard model.  We use this analysis 
in Sec.~\ref{sec:BoseHubbard} and
Sec.~\ref{sec:Disc}.
%This corresponds to the Bose-Hubbard model at unit filling and $U/t=3$. 

As argued in the main text, the correlation length of the MPS is expected to scale as $\chi^\kappa$ where $\kappa=6/(1+\sqrt{12})$~\cite{pollmann2009}.  For  distances that are smaller than this correlation length,  we expect  $\langle a^\dagger_ia_{j}\rangle\propto|i-j|^{-K/2}$.   Given a guess for the optimal Luttinger parameter, $K_0$,
we rescale:
\begin{align*}
    |i-j|\quad&\to\quad |i-j|~\chi^{-\kappa}, \\ 
    \langle a^\dagger_ia_j\rangle \quad&\to\quad \langle a^\dagger_ia_j\rangle~\chi^{\kappa K_0/2}.
\end{align*}
Following a procedure similar to Ref.~\cite{Kiely2022}, we then define an objective function which measures the deviation between the scaled density matrices with different values of $\chi$.  These should all collapse when $K_0=K$.  We adjust $K_0$ to minimize our objective function.  

%If the density matrix exhibits self-similar power-law scaling of the form described in Eq.~(\ref{eq:dMatLL}), then one will obtain a scaling collapse of all curves at the point $\eta=K/2$. 

% We %focus our attention on smaller separations while throwing out 
% discard
% values for $|i-j|<10$, as these exhibit strong non-universal features (see Fig.~\ref{fig:boseHubbard}). 
% %We also ignore separations where $|i-j|~\chi^{-\kappa}\gsim 0.05$, as these are in the  tails, where the scaling behavior breaks down.
% %For a trial $\eta$ we calculate 
% The simplest way to extract the

%Following a procedure which is similar to that in Ref.~\cite{Kiely2022}, we find the value of $\eta$ which minimizes the deviation between the rescaled density matrices, and
%
%
%Collecting all data for separations $|i-j|~\chi^{-\kappa}\lesssim 0.05$, we make linear fits for each value of $\eta$ and minimize 
%the sum of the squared deviations between the scaled curves. We take the value of $\eta$  that minimizes the squared deviations to be the optimal one, and we estimate the error by varying the data cutoffs by hand. With this technique, 
%we estimate the Luttinger parameter to be 

For the case in Sec.~\ref{sec:BoseHubbard}, where $\bar n=1$ and $U/t=3$, we find 
$K=0.423(2)$. Note that the value of the Luttinger parameter at the Mott lobe tip is 0.5, so this result is consistent with being on the superfluid side of the Mott-superfluid transition.

This same bond dimension scaling procedure is 
used to 
%applied to a variety of points in order to 
generate the data in Fig.~\ref{fig:scaling}.

\end{document}